\documentclass[iop]{emulateapj}
\usepackage{apjfonts}

\begin{document}

\epsscale{1.1}

\title{The WFC3 Galactic Bulge Treasury Program: Metallicity Estimates
for the Stellar Population and Exoplanet Hosts\altaffilmark{1}}
\shorttitle{The WFC3 Galactic Bulge Treasury Program}

\author{
Thomas M. Brown\altaffilmark{2}, 
Kailash Sahu\altaffilmark{2}, 
Jay Anderson\altaffilmark{2}, 
Jason Tumlinson\altaffilmark{2}, 
Jeff A. Valenti\altaffilmark{2}, 
Ed Smith\altaffilmark{2}, 
Elizabeth J. Jeffery\altaffilmark{2}, 
Alvio Renzini\altaffilmark{3}, 
Manuela Zoccali\altaffilmark{4}, 
Henry C. Ferguson\altaffilmark{2}, 
Don A. VandenBerg\altaffilmark{5}, 
Howard E. Bond\altaffilmark{2}, 
Stefano Casertano\altaffilmark{2}, 
Elena Valenti\altaffilmark{7}, 
Dante Minniti\altaffilmark{4,6}, 
Mario Livio\altaffilmark{2}, 
Nino Panagia\altaffilmark{2,8,9}
}

\altaffiltext{1}{Based on observations made with the NASA/ESA {\it Hubble
Space Telescope}, obtained at STScI, which
is operated by AURA, Inc., under NASA contract NAS 5-26555.}

\altaffiltext{2}{Space Telescope Science Institute, 3700 San Martin Drive,
Baltimore, MD 21218;  
tbrown@stsci.edu, 
ksahu@stsci.edu, 
jayander@stsci.edu, 
tumlinson@stsci.edu,
valenti@stsci.edu, 
edsmith@stsci.edu, 
jeffery@stsci.edu,
ferguson@stsci.edu,
bond@stsci.edu, 
stefano@stsci.edu, 
mlivio@stsci.edu, 
panagia@stsci.edu}

\altaffiltext{3}{Osservatorio Astronomico, Vicolo Dell'Osservatorio 5, 
I-35122 Padova, Italy; alvio.renzini@oapd.inaf.it}

\altaffiltext{4}{P. Universidad Cat$\acute{\rm o}$lica de Chile, 
Departmento de Astronom$\acute{\rm i}$a y Astrof$\acute{\rm i}$sica, 
Casilla 306, Santiago 22, Chile; mzoccali@astro.puc.cl, dante@astro.puc.cl}

\altaffiltext{5}{Department of Physics and Astronomy, 
University of Victoria, P.O. Box 3055, Victoria, BC, V8W 3P6, Canada; 
vandenbe@uvic.ca}

\altaffiltext{6}{Vatican Observatory, V-00120 Vatican City State, Italy}

\altaffiltext{7}{European Southern Observatory, 
Karl Schwarzschild\--Stra\ss e 2, D\--85748 Garching bei M\"{u}nchen, Germany;
evalenti@eso.org}

\altaffiltext{8}{INAF-CT, Osservatorio Astrofisico di Catania, 
Via S. Sofia 78, I-95123 Catania, Italy}

\altaffiltext{9}{Supernova Limited, OYV \#131, Northsound Road, 
Virgin Gorda, British Virgin Islands.}

\submitted{Accepted for publication in The Astrophysical Journal Letters}

\begin{abstract}

We present new UV-to-IR stellar photometry of four
low-extinction windows in the Galactic bulge, obtained with the Wide
Field Camera 3 on the {\it Hubble Space Telescope (HST)}.
Using our five bandpasses, we have defined reddening-free photometric
indices sensitive to stellar effective temperature and metallicity.
We find that the bulge populations resemble those formed via classical
dissipative collapse: each field is dominated by an old ($\sim$10~Gyr)
population exhibiting a wide metallicity range ($-1.5
\lesssim$~[Fe/H]~$\lesssim 0.5$).  We detect a metallicity gradient in
the bulge population, with the fraction of stars at super-solar
metallicities dropping from 41\% to 35\% over distances from the
Galactic center ranging from 0.3--1.2 kpc.  One field includes
candidate exoplanet hosts discovered in the SWEEPS {\it HST} transit
survey.  Our measurements for 11 of these hosts demonstrate that
exoplanets in the distinct bulge environment are preferentially found
around high-metallicity stars, as in the solar neighborhood,
supporting the view that planets form more readily in metal-rich environments.

\end{abstract}


\section{Introduction}

A primary quest of observational astronomy is elucidating the formation
history of massive galaxies. This work proceeds on two complementary
fronts.  High-redshift surveys investigate large galaxy samples over 
cosmic timescales, but do so with relatively crude age
diagnostics and limited spatial information in individual
galaxies.  Surveys of nearby galaxies investigate a small sample
in the present, but can probe discrete structures and
obtain diagnostics using photometry of individual stars.  The
Wide Field Camera 3 (WFC3; MacKenty et al.\ 2010) on the
{\it Hubble Space Telescope (HST)} is a powerful new tool to 
study galaxy formation in both the high-redshift and local regimes.

We know little about the formation of galaxy bulges, due in part to
conflicting evidence regarding our own bulge (see Zoccali 2010).
From a populations perspective, our bulge looks like a
``classical'' bulge -- similar to an old, nearly coeval
elliptical galaxy. From a morphological perspective, our bulge looks
like a ``pseudo-bulge,'' with a peanut shape apparently
arising from bar-driven secular processes. As summarized by Kormendy
\& Kennicutt (2004), a range of mechanisms may contribute to
bulge formation, with rapid processes in discrete events (e.g.,
dissipative collapse, mergers of clouds and proto-galaxies) dominating
in the early universe, and secular processes (interactions
between stars, gas clouds, bars, spiral structure, triaxial halos,
etc.) dominating at later times.  In broad terms, we see classical
bulges if the earlier rapid processes are dominant, and pseudo-bulges
if the later slower processes are dominant.  However, a new paradigm
has recently emerged, driven by observations at $z \sim 2$,
that reveal the widespread existence of large, rotating, disk galaxies
with much higher gas fractions compared to local spirals (Genzel et
al.\ 2008; F\"orster Schreiber et al.\ 2009).  Such gas-rich, clumpy
disks are prone to instabilities that can lead to bulge formation at
early cosmic epochs and over timescales much shorter than those
traditionally associated with secular instabilities in local spirals
(e.g., Immeli et al.\ 2004; Elmegreen et al.\ 2009).  These processes
should imprint distinct age and metallicity gradients upon the
bulge stellar populations.  For example, bulge evolution dominated by
secular processes (such as bar instabilities) is not expected to
produce radial metallicity gradients, but is expected to
produce younger stars in the bulge outskirts (see Zoccali 2010 and
references therein).

To shed light on these processes, our WFC3 Galactic Bulge Treasury
program uses the new panchromatic capabilities of {\it HST} to probe
the bulge stellar populations as a function of position (Brown et
al.\ 2009).  One of these bulge fields includes 13 candidate exoplanet
hosts discovered via the SWEEPS transit survey (Sahu et
al.\ 2006), and so our photometry can also demonstrate where these
hosts fall within the bulge metallicity distribution function (MDF).
Because WFC3 is a relatively new addition to {\it HST},
the full analysis of these data will require
improvements in the calibration pipeline, data reduction, and analysis
techniques, but we present here a 
look at preliminary results from our program.

\section{Observations and Data Reduction}

Our program is summarized in Brown et al.\ (2009), but we give a brief
summary here.  We obtained deep photometry in four
low-extinction bulge windows (Table 1) with {\it HST}/WFC3.  To
measure proper motions, we will repeat observations in three fields,
while the fourth (SWEEPS) was observed
previously with {\it HST} by Sahu et al.\ (2006).  WFC3 has two
imaging channels: a UVIS channel comprised of two 4k$\times$2k
UV/optical CCDs, and an IR channel employing a single 1k$\times$1k
HgCdTe array.  We observed these bulge fields with five channel/filter
combinations: UVIS/F390W (hereafter $C$; exposure time 11180 s), 
UVIS/F555W ($V$; 2283 s), UVIS/F814W ($I$; 2143 s), IR/F110W ($J$; 1255 s), and
IR/F160W ($H$; 1638 s).  The bulge images were dithered to allow
2$\times$2 resampling of the point spread function (PSF);
because the IR channel has a smaller field of view ($123 \times
136^{\prime\prime}$) than the UVIS channel ($162 \times
162^{\prime\prime}$), additional offsets provided IR coverage of the full
UVIS image area.  As part of this program, we also observed six
star clusters in the same filters, although with less
exposure time and little dithering; the resulting color-magnitude diagrams
(CMDs) are much noisier than the bulge CMDs, but sufficient to determine
ridge lines and calibrate isochrone transformations.  The first epoch
of observations spans 2009 Oct to 2010 Sep.

\begin{table}[t]
\begin{center}
\caption{Bulge Fields}
\begin{tabular}{lllllr}
\tableline
                & $l$      & $b$    & $R_{min}$\tablenotemark{a}  & $A_V$   & $A_V$ \\
Name            & (deg)    & (deg)  & (kpc) & (mag) & Reference \\
\tableline
Stanek's Window & +0.25  & -2.15 & 0.32  & 2.6  & Stanek (1998) \\
SWEEPS\tablenotemark{b}  & +1.26  & -2.65 & 0.43  & 2.0  & Sahu et al. (2006) \\
Baade's Window  & +1.06  & -3.81 & 0.58  & 1.6  & Baade (1963)\\
OGLE29\tablenotemark{c}  & -6.75  & -4.72 & 1.21  & 1.5  & Sumi (2004) \\
\tableline
\end{tabular}
\end{center}
\tablenotetext{a}
{Projected Galactocentric distance, assuming a distance of 8.4 kpc
(Reid et al.\ 2009).}
\tablenotetext{b}
{Sagittarius Window Eclipsing Extrasolar Planet Survey}
\tablenotetext{c}
{Optical Gravitational Lensing Experiment}
\end{table}

The pipeline currently applies a mix of calibrations from
ground and preliminary in-flight measurements.  
We registered, geometrically corrected, and combined the calibrated
images using {\sc drizzle} (Fruchter \& Hook 2002), with rejection of
hot pixels and cosmic rays for the UVIS images (cosmic rays are
removed during the pipeline processing of the non-destructive IR
readouts). Using {\sc daophot-ii} (Stetson 1987), we performed
PSF-fitting photometry in the UVIS images and aperture photometry
(with neighboring PSFs subtracted) in the IR images.  The
photometry was corrected to the Vegamag zeropoints of Kalirai et 
al.\ (2009a, 2009b), using relatively isolated stars.  CMDs for the 
main-sequence (MS) population in each field are shown in Figure 1.

It is worth noting some limitations of the current reduction.  The
flat-field corrections are based upon ground tests, with residual
spatial variations in sensitivity at the $\sim$5\% level.  The
geometric distortion correction is undergoing significant
revisions, limiting image registration accuracy.
The photometric zeropoints were calculated
with a subset of the in-flight calibration standards.
Some of the UVIS exposure times are short, to
provide giant star photometry in each field, but with photometric
accuracy currently limited by uncertainties in the CCD charge 
transfer efficiency. The IR
detectors exhibit persistence, elevating the dark
current where a bright source has recently illuminated the detector;
a future persistence correction will reduce the photometric noise
for the faint stars.

\begin{figure*}[t]
\plotone{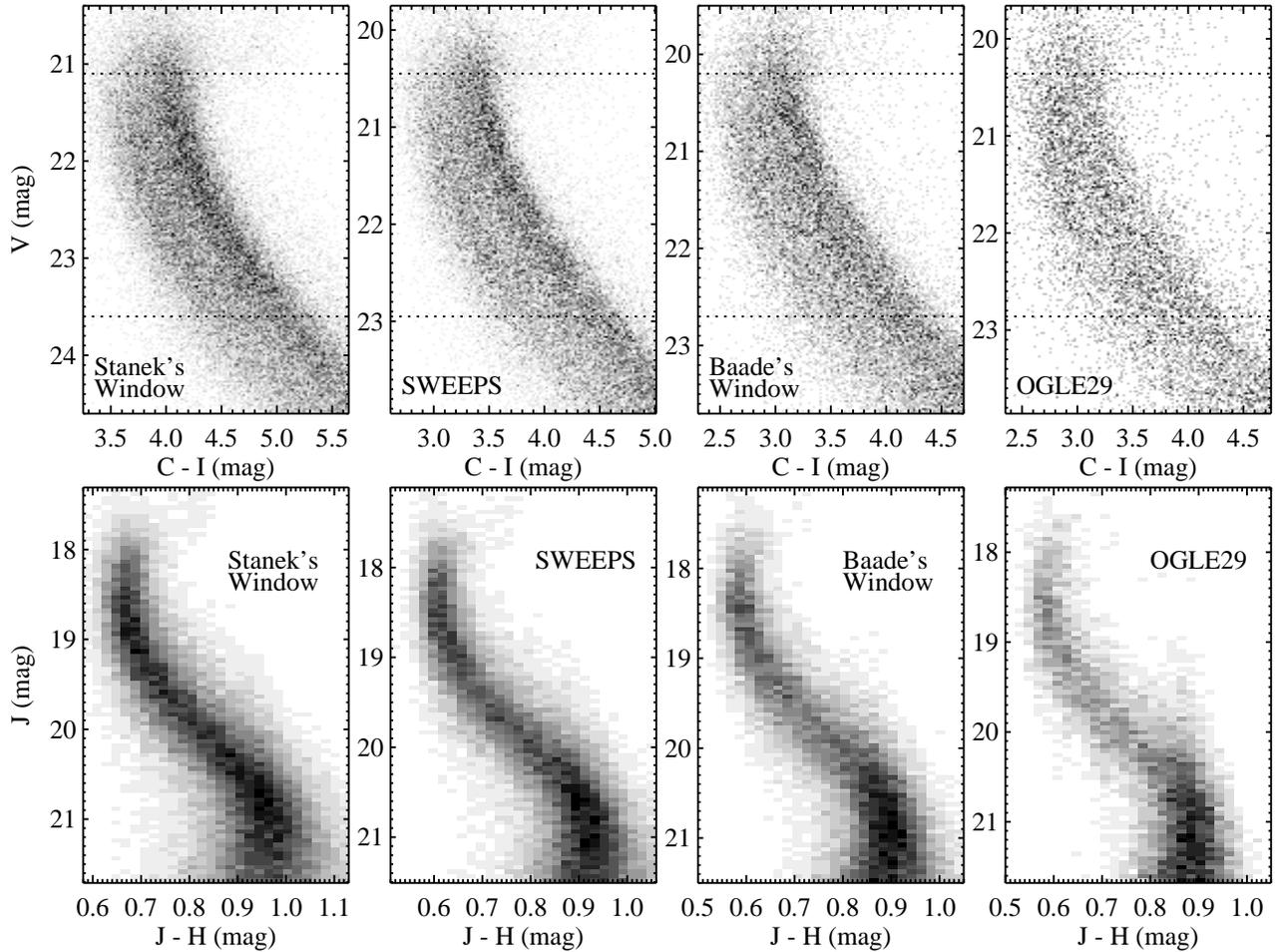}
\caption{CMDs ({\it grey shading}) for our bulge fields,
  binned into Hess diagrams with a
  linear stretch.  The dashed lines in the top panels
  isolate stars on the upper 2.5~mag of the MS.
  In the inner fields (Stanek's Window and
  SWEEPS), the $C - I$ color distribution on the MS is
  bimodal, with a strong red ridge and a fainter blue ridge.  Our 
  metallicity index indicates that the red ridge arises from the
  large number of metal-rich stars, while the blue ridge
  is due to color degeneracy at low metallicity.  CMDs 
  constructed from bands at wavelengths longer than $C$
  (e.g., $V$ and $I$ in Sahu et al.\ 2006) do not show this bimodality 
  because they are much less sensitive to metallicity.}
\end{figure*}

\section{Analysis}

\subsection{Population Age}

The IR CMD of low-mass MS stars exhibits a ``knee'' at
$\sim$0.5~$M_\odot$, due to collisionally-induced 
H$_{\rm 2}$ absorption, which serves as a standard candle (Bono et
al.\ 2010).  Our IR photometry reaches stars well below this knee
(Figure 1).  The magnitude difference between the
knee and the MS turnoff (MSTO) is approximately constant across
our fields, indicating a predominantly old ($\sim$10~Gyr) population
in each field.  The characterization of any
subtle variations in the age distribution with position in the bulge
will require further analysis.

\begin{figure*}[t]
\plotone{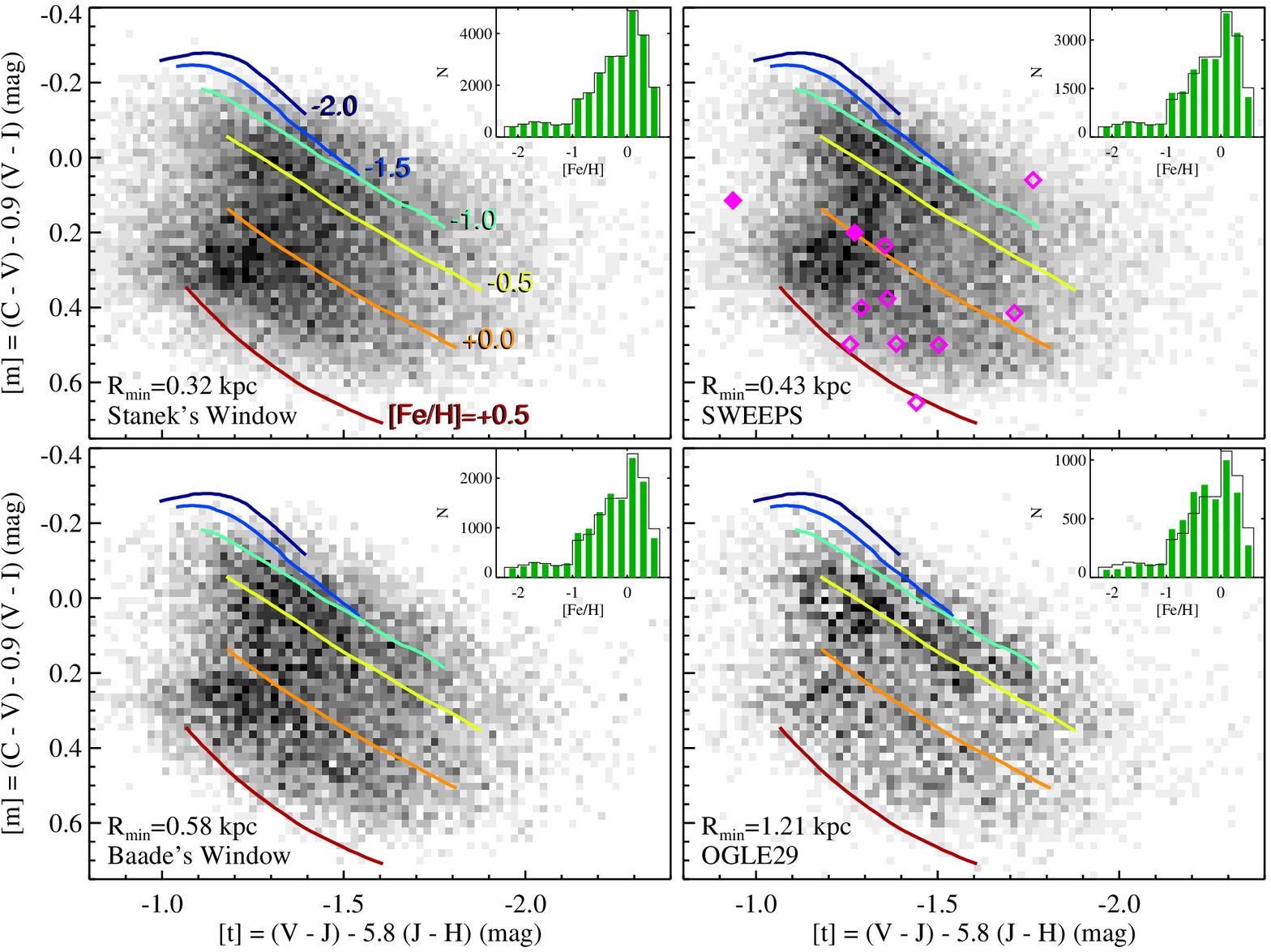}
\caption{Broad-band indices of temperature and metallicity for
  stars on the upper 2.5~mag of the MS in each bulge
  field ({\it grey shading}).  For a given extinction law,
  these indices are insensitive to the amount of reddening, but the
  photometry here has been corrected for field-to-field variations in
  reddening law (see \S3.2).  The curves represent the upper MS in 10~Gyr
  isochrones, ranging in [Fe/H] from $-2.0$ to $+0.5$ 
  ({\it labeled}).  In the SWEEPS field, candidate exoplanet
  hosts are indicated (Sahu et al.\ 2006; {\it diamonds}),
  clustering at high metallicity.  Radial velocities
  support the planetary nature of two exoplanet candidates
  ({\it filled diamonds}).  The inset for each panel shows
  the estimated MDF in each field ({\it
  green bars}), along with that in
  Stanek's Window for comparison ({\it black histogram}; 
  normalized to the counts in each field).  Note that the
  metallicity estimates are degenerate at low metallicity,
  artificially producing a long metal-poor tail in the MDF.
  The general tendency is for decreasing metallicity at
  increasing distance from the Galactic center.}
\end{figure*}

\subsection{Anomalous Extinction}

As demonstrated by Brown et al.\ (2009), our five photometric bands
can be combined to provide reddening-free indices of
temperature [$t$] and metallicity [$m$] in the stellar population.
Specifically, for a given extinction law (parameterized by $R_V \equiv
A_V / E(B-V)$), the [Fe/H] inferred from these indices varies by only
$\sim$0.1 dex when $A_V$ varies by a magnitude (see Figures 3 and
4 of Brown et al.\ 2009).  For old MS stars reddened with a
Galactic average extinction law (Fitzpatrick 1999; $R_V = 3.1$), these
indices can be defined as:

\begin{equation}
 [t] \equiv (V - J) - 5.8 \times (J - H)
\end{equation}
and
\begin{equation}
[m] \equiv (C - V) - 0.90 \times (V - I).
\end{equation}

However, significant deviations from the Galactic average extinction 
law will shift the distribution in the [$m$] vs. [$t$]
plane, because the coefficients in these indices depend upon the
reddening vectors in each color.  Such shifts are expected here,
because Sumi (2004) has found significant variations in $R_I \equiv A_V /
E(V-I)$ toward Galactic bulge fields.  The large coefficient (5.8) in
the [$t$] index makes it particularly sensitive to random and
systematic errors in $J-H$.  To facilitate the comparison of the
populations in each field, we applied a correction for this variation
in extinction law before calculating the [$m$] and [$t$] indices.  We
derived these color shifts by comparing the
MS loci of the bulge populations in CMDs of $C$ vs. $C-V$,
$V$ vs. $V-I$, $I$ vs. $I-J$, and $J$ vs. $J-H$.  Although the color
distribution within the MS locus varies from field to
field, the characteristic shape and boundaries of the MS
locus in each field exhibit little variation, such that our shifts
allow the loci to be aligned in color to an accuracy of
$\sim$0.01~mag.  To this end, we aligned the MS locus in
each field to that in Baade's Window.  The implied extinction curve,
when compared to the Galactic average ($R_V = 3.1$)
with both normalized at the red end ($H$), differs by less than 0.1~mag
in any band, and by less than 0.02~mag if $2.8 < R_V < 3.1$.
Note that our extinction law corrections are completely empirical;
they are consistent with a somewhat lower $R_V$ at decreasing distance
from the Galactic center, but do not
allow us to derive the actual value of $R_V$ at an
interesting level of confidence.  We assume that the extinction law
varies little within the relatively small {\it HST} field of view.

\subsection{Metallicity Distributions in the Bulge Populations}

Once the variation in extinction law is corrected (see \S3.2), we can
compare the MDFs in each field.  To restrict the
analysis to a relatively narrow temperature range where the indices
are well-behaved, we selected stars between the
MSTO and a point 2.5~mag fainter in $V$ (Figure 1; {\it
dashed lines}).  These relatively bright stars are also well above
the detection limits in each field, with small photometric errors due
to photon counts and stellar crowding.

The distributions in the [$m$] vs. [$t$] plane for each field are
shown in Figure 2.  For comparison, we show the expectations from a
set of 10~Gyr isochrones spanning a wide range in [Fe/H]
(VandenBerg et al.\ 2006), transformed into the [$m$] vs. [$t$] plane
using the NextGen grid of synthetic spectra (Hauschildt et al.\ 1999;
updated).  We note that new evolutionary tracks and isochrones, 
accounting for helium diffusion and recent advances in basic stellar
physics, are being computed for application to
these data (by D.A.V.).  As with the photometry, the isochrones are shown 
from the MSTO to a point 2.5~mag fainter in $V$. 
Spectroscopy of the bulge population indicates significant enhancement
in [$\alpha$/Fe] at low metallicity, trending to less enhancement
as one approaches [Fe/H]$=0$ (Lecureur et al.\ 2007).  To mimic that trend,
our chosen isochrones assume [$\alpha$/Fe]=0.3 for [Fe/H]$\leq 0$ and
[$\alpha$/Fe]=0.0 for [Fe/H]=0.5.  The synthetic spectra assume the
same $\alpha$-enhancement, but are interpolated from
spectra with [$\alpha$/Fe] values of 0, 0.2, and 0.4.  Loci of
constant metallicity form approximately parallel lines in the diagram,
with [$m$]~=~[$m_0$]~$-0.6\times$[$t$], such that interpolation
between these lines provides a metallicity estimate for each star.  
The derived MDFs are shown in Figure 2 ({\it histogram insets}).  In Baade's 
Window, our MDF at [Fe/H]$> -0.5$ is similar to that of Zoccali et al.\ (2008; 
from spectroscopy of giants) but has more stars at lower metallicities.

The inner fields (Stanek's Window and SWEEPS) exhibit a
distinctly bimodal distribution of color in the $V$ vs. $C-I$ CMDs
(Figure 1), while the outer fields (Baade's Window and OGLE29)
exhibit a bimodal distribution in the [$m$] vs. [$t$] plane
(Figure 2), which is reflected in the inferred MDFs.
This is due to the nonlinear relationship between the
indices, traditional CMD colors, and chemical composition.  If the
position in the [$m$] vs. [$t$] plane is taken as an indication of
metallicity, the stars at [Fe/H]$>0$ overwhelmingly lie in
the prominent red ridge of the $V$ vs. $C-I$ CMDs, while the stars
at [Fe/H]$<-1$ tend to lie in the weaker blue ridge of these CMDs.

Brown et al.\ (2008) discussed the advantages of the [$m$] vs. [$t$]
diagram regarding depth effects, but made no mention of binaries.  In
a CMD, depth blurs the photometric distribution along the ordinate,
blurring the inferred MDF when comparing to isochrones, but a
color-color diagram is insensitive to depth.  In a CMD, binaries are
brighter and redder than the single-star sequence; comparison of
binaries to isochrones can thus overestimate their metallicities.  In
our color-color diagram, a companion of either equal or much lower
mass will not shift the position of the primary, but a companion that
is only somewhat fainter and cooler will artificially shift the
implied metallicity lower (i.e., opposite to the CMD effect).  This is
because the secondary has a larger impact on [$t$] than [$m$].  The
metallicities of any binaries in our field can be underestimated by
up to $\sim$0.3~dex.

From a strictly statistical perspective, a Kolmogorov-Smirnov (KS) test
indicates that no pair of these fields has populations drawn from the
same distribution; the chance is 3\% for the populations in the SWEEPS
field and Baade's Window, and $\ll$1\% for any other pair.
The MDFs in the two innermost fields (Stanek's
Window and SWEEPS) are similar in appearance, although the metallicity
appears to be shifted to slightly lower [Fe/H] in the SWEEPS field.
Progressing outward, Baade's Window and the OGLE29 field each exhibit
more stars shifting to lower metallicities, with the MDFs
appearing more bimodal than that in the interior fields,
due to a stronger presence of relatively metal-poor stars.  Taking the
relationship between indices and [Fe/H] at face value, the fraction of
stars with super-solar metallicities drops as one progresses outward
through the fields: Stanek's Window (41\%), SWEEPS (39\%), Baade's
Window (38\%), and OGLE29 (35\%).  Rich et al.\ (2007) found no
gradient in the bulge metallicity when comparing the spectroscopic
metallicities of 17 M giants in the inner bulge to 14 M giants in
Baade's window, but given their small sample, their result may be
consistent with our findings.  Zoccali et al.\ (2008) found the bulge
to decrease in metallicity along the minor axis beyond 4$^{\rm o}$.

Note that our measured MDF variation is for
the general population in each field, and cannot be investigated for
the bulge in isolation without proper-motion cleaning of
the foreground stars.  We have used the TRILEGAL Galaxy model,
with its default parameters (Girardi et al.\ 2005), to estimate that the 
foreground thin disk, thick disk, and halo together contribute 5--9\% 
of the total population in each field, with higher contamination at
increasing distance from the Galactic center.  We performed a
preliminary proper-motion cleaning of the catalog in the SWEEPS field,
by comparing our astrometry to that in Sahu et al.\ (2006), and
assuming the relative disk and bulge velocities of Clarkson et
al.\ (2008).  This cleaning clearly reduces the presence of CMD
features associated with the disk foreground (e.g., the blue plume
above the old MSTO), but the MDF
in the remaining bulge population is not significantly changed from
that shown in Figure 2.

The combination of isochrones and synthetic spectra employed in Figure
2 was chosen because it spans nearly the full range of photometric
indices in each field, demonstrating the nonlinear relationship
between indices and [Fe/H].  However, the zeropoint of the [Fe/H]
scale is uncertain at the level of 0.2--0.3 dex, depending upon the
actual extinction law in Baade's Window (changing $R_V$ by 0.1 shifts the
implied [Fe/H] by 0.1 dex), the assumed abundance pattern (e.g.,
[$\alpha$/Fe] as a function of [Fe/H], particularly at high [Fe/H]),
the photometric zeropoints, and the spectral library employed
in the isochrone transformation.
Putting the empirical ridge lines of our star clusters (\S2) in the same
[$m$] vs. [$t$] plane would imply our assumed [Fe/H] scale
underestimates the true [Fe/H] by $\sim$0.2--0.3 dex down to
[Fe/H]~$\approx -1.5$, and that our metallicity scale is degenerate at
lower metallicities.  The offset may be a real systematic error in our
metallicity scale, but it may also be due to distinctions in the
reddening law and/or abundance pattern between the cluster and bulge
populations.  The cluster photometry was not corrected for variations
in extinction law in the same manner used for the bulge fields,
because the MS locus for each cluster is very distinct from
that in the bulge, given the single metallicity yet noisier
photometry.

\subsection{Metallicities of the Candidate Exoplanet Hosts}

Our SWEEPS field includes 13 of the 16 candidate exoplanet hosts found
in the {\it HST} transit survey of Sahu et al.\ (2006).  Two of
these 13 candidates are too faint and red to appear in our $C$ images,
but the remaining 11 have photometry enabling their placement in
the [$m$] vs. [$t$] plane for the general SWEEPS population (Figure 2;
{\it diamonds}).  Two of these 11 candidates have radial velocities
that support their planetary nature, and are highlighted
in Figure 2 ({\it filled diamonds}).  Although the zeropoint
for our assumed [Fe/H] scale is uncertain at the level of
$\sim$0.2--0.3 dex (see \S3.3), the relative [Fe/H] measurements for
the exoplanet hosts and the general population in the SWEEPS field are
much more secure, because the same systematic uncertainties apply to
both.  It is clear from Figure 2 that the candidate exoplanet hosts
predominantly fall in the metal-rich end of the bulge MDF -- a population 
that is already
skewed toward high metallicity.  Aside from a single candidate at the
metal-poor end of the distribution, the remaining 10 candidates are
more metal-rich than half the population, with 7 in the top quartile.
A KS test of the implied metallicities in the
exoplanet hosts and general population indicates that the chance they
are both drawn from the same parent population is less than 2\%. \\ \\ \\

\section{Discussion}

We have performed a preliminary analysis of the data from our WFC3
Galactic Bulge Treasury Program.  Keeping in mind the various
systematic uncertainties at this stage, our analysis of the dwarf
stars supports the picture of the bulge gleaned from investigations of
the brighter giant stars (see Zoccali 2010 and references therein).
Our IR photometry reaches the knee on the lower MS, and
indicates that the population is predominantly old ($\sim$10~Gyr) in
all of the bulge fields, with no obvious age gradient.  The declining
metallicities at increasing radius are seemingly inconsistent with the
secular processes that are traditionally associated with the
formation of a peanut-shaped bulge.  Our findings are consistent with
a classical bulge formed via rapid dissipative collapse
(either monolithic or via the merger of independent components), but also
consistent with a recently emerging formation paradigm, motivated by
observations of gas-rich spirals at $z \sim 2$ (Genzel et al.\ 2008;
F\"orster Schreiber et al.\ 2009).  In this new paradigm,
instabilities in gas-rich disks can drive early bulge formation over
rapid timescales (e.g., Immeli et al.\ 2004; Elmegreen et
al.\ 2009).

Of the hundreds of extrasolar planets discovered to date, most have
been found in the solar neighborhood via radial-velocity measurements.
A notable exception is the discovery of 16 candidate exoplanet hosts
in the SWEEPS transit survey of the Galactic bulge (Sahu et
al.\ 2006).  Our multiband photometry of 11 of these hosts
demonstrates that they fall almost exclusively at the high end of the
MDF in this high-density, metal-rich field.
Exoplanets in the distinct environment of the solar neighborhood are
also found preferentially at high metallicity (e.g., Fischer \&
Valenti 2005).  Out of the $\sim$500 exoplanets discovered to date
whose orbital periods range from a fraction of a day to well over five
years, $>100$ have orbital periods less than five days, implying
significant migration since formation.  The correlation of such
planets with stars of high metallicity probably indicates that planets
are preferentially formed in high-metallicity environments, or 
alternatively that planets migrate more easily under such conditions.

\acknowledgements

Support for Program 11664 was provided by NASA through a grant
from STScI, which is operated by AURA, Inc., under NASA contract NAS
5-26555.  MZ acknowledges Fondecyt Regular 1085278.  AR acknowledges
ASI for support via the grant ``COFIS-Analisi Dati.'' We appreciate
useful discussions with J. Kalirai and A. Dotter.

\end{document}